\documentclass[10pt,journal,compsoc]{IEEEtran}
\usepackage{graphicx,subfigure}
\graphicspath{{figures/}}
\usepackage{url}
\usepackage{color,soul}
\usepackage{soul}

\ifCLASSOPTIONcompsoc
  \usepackage[nocompress]{cite}
\else
  \usepackage{cite}
\fi
\ifCLASSINFOpdf
\else
\fi
\begin{document}
%
\title{An Ambient-Physical System to Infer Concentration in Open-plan Workplace}
%
%
%
%

\author{Mohammad~Saiedur~Rahaman,~\IEEEmembership{Member,~IEEE,}
        Jonathan Liono,~\IEEEmembership{Member,~IEEE,}
        Yongli~Ren,~\IEEEmembership{Member,~IEEE,}
        Jeffrey~Chan,~\IEEEmembership{Member,~IEEE,}
        Shaw~Kudo,
        Tim~Rawling, 
        and Flora~D.~Salim,~\IEEEmembership{Member,~IEEE}

\IEEEcompsocitemizethanks{\IEEEcompsocthanksitem M. S. Rahaman, J. Liono, Y.~Ren, J.~Chan and F.~D.~Salim are with Computer Science and Information Technology, School of Science, RMIT University, Melbourne, VIC 3000, Australia.\protect\\
E-mail: \{saiedur.rahaman, jonathan.liono, yongli.ren, jeffrey.chan, flora.salim\}@rmit.edu.au
\IEEEcompsocthanksitem S.~Kudo and T.~Rawling are with Arup, Melbourne, VIC, Australia.\protect\\
E-mail: \{shaw.kudo, tim.rawling\}@arup.com}

}

\IEEEtitleabstractindextext{%
\begin{abstract}
One of the core challenges in open-plan workspaces is to ensure a good level of concentration for the workers while performing their tasks. Hence, being able to infer concentration levels of workers will allow building designers, managers, and workers to estimate what effect different open-plan layouts will have and to find an optimal one. In this research, we present an ambient-physical system to investigate the concentration inference problem. Specifically, we deploy a series of pervasive sensors to capture various ambient and physical signals related to perceived concentration at work. The practicality of our system has been tested on two large open-plan workplaces with different designs and layouts. The empirical results highlight promising applications of pervasive sensing in occupational concentration inference, which can be adopted to enhance the capabilities of modern workplaces.
\end{abstract}

}

\maketitle

\IEEEdisplaynontitleabstractindextext

%
\IEEEpeerreviewmaketitle

\IEEEraisesectionheading{\section{Introduction}\label{sec:introduction}}

%
%
%
%
\IEEEPARstart{O}ver the past decade, there has been an upward trend for the adoption of open-plan offices \cite{Montanari:2017:DEA:3139486.3130951,doi:10.1080/00140139.2016.1188220,doi:10.1177/0013916582145002} in the corporate sector. This trend shift is often associated with a drive to save space, reduce costs and accommodate growing teams. The shared workplace has been shown to bring greater worker satisfaction \cite{doi:10.1080/17508975.2012.695950,eric:academy,KIM201318} and provide work flexibility \cite{doi:10.2307/3556639,doi:10.1177/0018726709342932}. Occupants in a shared workplace have the potential to exchange knowledge more effectively \cite{Meijer:TF,doi:10.1002/job.1973,KIM201318} which can enrich their skills and potentially lead to greater productivity \cite{doi:10.1080/17508975.2012.695950} while performing their work-related tasks. However, at the same time, studies have identified a number of problems \cite{10.1371/journal.pone.0193878,MORRISON2017103} associated with open-plan offices including increased distrust, distractions, and uncooperative behavior. In this paper, our study focuses on the exploration of various factors that influences the worker concentration (i.e. ability to focus on the task at hand while ignoring distractions) in an open-plan workplace setting using pervasive sensing. Concentration has been modelled in a real-time manner. Outcomes of this study can help designers better plan modern workplace layouts and allow ongoing support during workplace occupation to provide personalized recommendations for workers using the space. Further, pervasive sensing for concentration inference can help identify the key drivers of perceived concentration levels among workers, ultimately leading to workplace designs that more carefully consider the commonly identified problems of open-plan offices.

\begin{figure}[t!]
  \centering
  \includegraphics[width=2.3in]{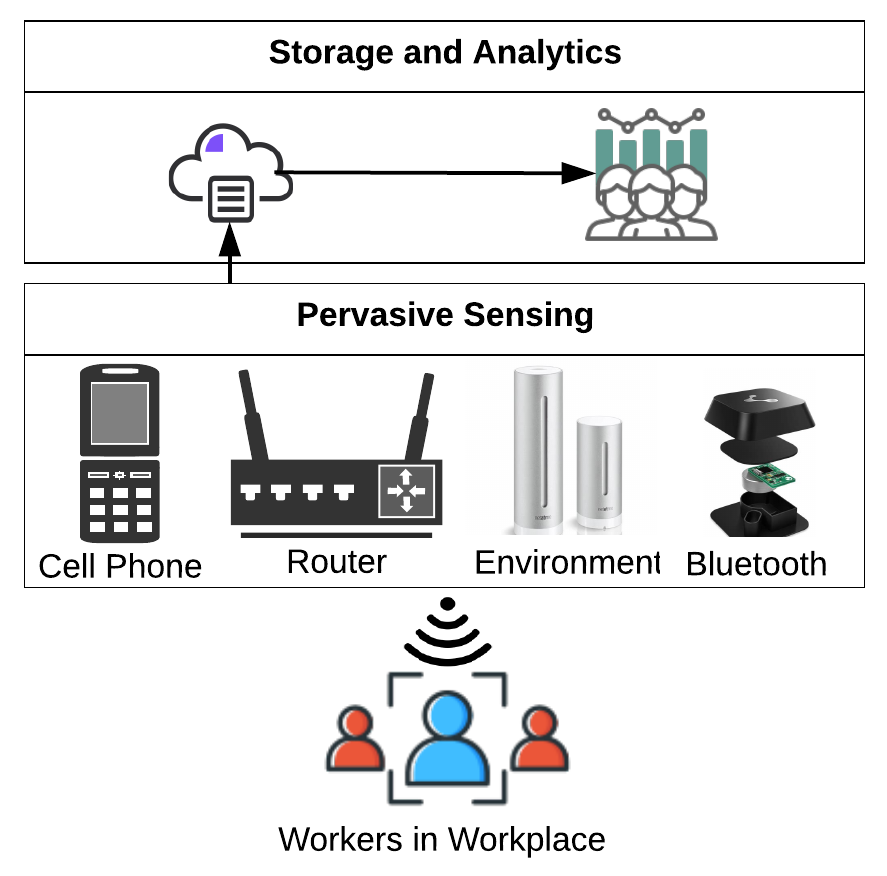}
  \caption{Overview of pervasive sensing and analytics in workplace.}
  \label{fig:Pervasive-sensing-at-work}
\end{figure}

Recent technological developments and the proliferation of pervasive technologies has opened up many opportunities to collect data from various sensors and smart devices. These diverse datapoints can be fused together to produce intelligent analytics. Ultimately, this enables informed decision making at a more granular level. In the context of open-plan workplaces, data sensed actively and passively using pervasive devices can be used for monitoring overall organizational behavior \cite{4694078} and workplace interactions of employees \cite{Brown:2014:TSI:2531602.2531641,GHAHRAMANI201842,Mark:2014:CMF:2531602.2531673,Brown:2014:AIT:2632048.2632056, Mashhadi:2016:CSC:2957265.2957272}. These outcomes can help improve the understanding of the emotions, team dynamics and productivity of employees, which are all key factors for organizational success. Figure \ref{fig:Pervasive-sensing-at-work} illustrates the overall conceptual architecture of using pervasive sensors in workplaces. While research of pervasive sensing in work environments has been popular in recent years, there has been little research done on using intelligent sensing applications to infer concentration in workplace environments.

The core challenge of using pervasive sensing for concentration inference in open-plan workplaces is the dynamic concentration traits of workers. Therefore, building a generalized model is challenging. In this research, we conduct a novel investigation of the problem of concentration inference using data collected from pervasive sensors. We develop a system that captures, integrates, models, and provides quantitative analysis and semantic abstraction of key factors that relates to the concentration in the workplace. We specifically investigate how to measure workers' concentration from their smartphone sensors (e.g. sedentary movements) and ambient sensors (e.g. indoor noise, temperature, humidity, and CO\textsubscript{2} density). We present the task of concentration inference as a prediction problem. We show that the physical (i.e. worker movements) and ambient factors together can improve the predictability of workers' concentration. We validate the performance and applicability of our concentration inference system against the self-annotated concentration levels collected from the participants (corporate workers) of two pilot sites in one of the major Australian cities. The key contribution of this paper includes:

\begin{itemize}

    \item A novel system that leverages pervasive sensing to capture and integrate concentration related factors, and infer concentration levels of workers in an open-plan work environment. 
    
    \item Investigation of key factors influencing concentration at work through quantitative analysis and semantic abstraction.
    
    \item Implication and design considerations of concentration inference using pervasive sensing for  enabling  future  capabilities  of  ubiquitous  computing  applications in smart buildings/spaces.
    
\end{itemize}

The remainder of the paper is structured as follows. In Section~\ref{sec:related-work}, we describe works related to concentration study and measurement. Section~\ref{sec-problem} describes the problem formulation. Section~\ref{sec:proposed-framework} describes our proposed ambient-physical inference system. Section~\ref{sec-experiment} describes the experimental results along with the implications and design consideration of concentration inference system at open-plan workplace. The paper concludes in Section~\ref{sec-conclusion} with a summary of our findings and potential avenues for future work.

\section{Concentration and Measurement}
\label{sec:related-work}

This section provides an overview of the current state of concentration study and measurement. Current research mainly focuses on the need for concentration in the workplace in relation to different office design or layout. The main focus of this has been on concentration needs and associated factors in many domains including environmental psychology, ergonomics and corporate real estates \cite{SEDDIGH2014167,doi:10.1080/00140130903389019,doi:10.1080/15487768.2016.1162758}.

Research has identified the high need for concentration in open-plan offices \cite{SEDDIGH2014167}. A recent study has found that noise and lack of privacy are two main reasons for concentration difficulties in open-plan work environments \cite{doi:10.1080/00140130903154579,pejtersen_poulsen_2006}. Another study concluded that participants remember fewer things (i.e. words) in a noisy surrounding compared to a low noise environment which can also negatively impact their motivation to work \cite{JAHNCKE2011373}. An investigation on cognitive performance impairment caused by irrelevant background speech with varying degrees in open-plan office has been conducted by \cite{JAHNCKE2013307}, which found that a strong relationship between speech intelligibility and the cognitive performance of workers. The research also identified that several office-tasks are more prone to distraction than others. A large number of participants from another study consisting of 88 samples found that concentration of employees can be impaired by various office noise including telephones left ringing at vacant desks and background noise from people talking \cite{doi:10.1080/00140130412331311390}. Correlation between stress and concentration has been identified in studies noting that high stress level is one of the most significant sign of low concentration and and poor decision making \cite{hse2007}.

In recent years, further research has been conducted by using wearables to monitor workers inside their office spaces. The use of wearable technologies paired with machine learning and data analytics has been applied to measure physiological parameters in the workplace \cite{doi:10.1111/isj.12205}. Pervasive signals from the wearable sensors were used to identify various types of interactions in the workplace \cite{Mashhadi:2016:CSC:2957265.2957272}. The combination of physiological and sociometric sensors are used to identify stress of people in a social situation using machine learning techniques \cite{mozos_2017}. A similar study used wearable physiological sensors to identify stressful and non-stressful situations \cite{10.1007/978-3-319-18914-7_55}. A recent study utilized data collected from wearable devices to investigate team dynamics in relation to space usage and organizational hierarchy \cite{Montanari:2017:DEA:3139486.3130951}. The use of mobile sensor data has gained popularity in recent years to measure social interaction among people \cite{5959586}. Another research by analyzing indoor co-location and instant messenger data for automatic inference of social relationship among people in an organization is presented in \cite{Choi:2013:MSR:2441776.2441811}.

Despite various studies, concentration studies considering workers in an office setting have mainly been focused on the feedback and perceptions of the workers about their workplace environment. Yet, no research has been found that uses pervasive sensing passively to measure workers concentration levels while they are at work. The automatic inference of concentration at work from passively collected data has the potential not only to help inform management and workers of trends and factors but also to prevent the negative outcomes associated with chronic concentration disorder. This research presents an ambient-physical system to infer concentration in an open-plan workplace using pervasive sensing.

\section{Problem Formulation}
\label{sec-problem}

In this section, we define concentration, ambient and physical features, and formulate the problem of concentration inference using pervasive sensing of ambient-physical features.

\textbf{Definition 1:} \textit{Concentration} is the level of attentiveness in the workplace. We use a five-point Likert scale to measure the perceived concentration levels of worker participants where 1-point, 2-point, 3-point, 4-point, and 5-point indicate very poor, poor, neutral, high, and very high concentration levels respectively.

\textbf{Definition 2:} \textit{Ambient features} capture the factors related to surrounding environment that affect concentration in the workplace. The factors include noise levels, temperature, indoor air quality (i.e. CO\textsubscript{2} concentration), humidity, air pressure, and surrounding electromagnetic field. There may have other factors that can affect a person's concentration levels. Note that the aim of this research is not to identify and investigate all the ambient factors, rather we show how ambient features can help infer concentration levels of workers in the workplace.

\textbf{Definition 3:} \textit{Physical features} capture the movements of workers in the workplace. We show that the physical features are also a good indicator for many workers in the workplace. These features are devised from the reading of different mobile sensors, including accelerometer, gyroscope, and pedometer.

\textbf{Concentration inference:} We formulate the concentration inference task as a classification problem. Let, $F_a$ and $F_p$ denote the sets of ambient and physical features of concentration level associated with the $i^{th}$ worker, $u_i$. In our concentration dataset, each instance is described by a timestamped vector $<u_i, F_a, F_p, \mathcal{L}_c>$ where, $\mathcal{L}_c$ is one of the five different concentration levels.

Given a set of training samples, we train a number of classification algorithms to infer the corresponding concentration label of a new instance. Note that the user identification, $u_i$ was removed during this training phase of classifiers. The concentration inference task can be formally represented as:

\begin{equation}
    f(F_{a}, F_{p}) \rightarrow \mathcal{L}_c
\end{equation}
where $f(\cdot)$ is a function that establishes a mapping between ambient-physical features and the perceived concentration level, $\mathcal{L}_c$ of a worker.

\section{Proposed Ambient-Physical System}
\label{sec:proposed-framework}
\begin{figure}[h]
  \centering
  \includegraphics[width=\columnwidth]{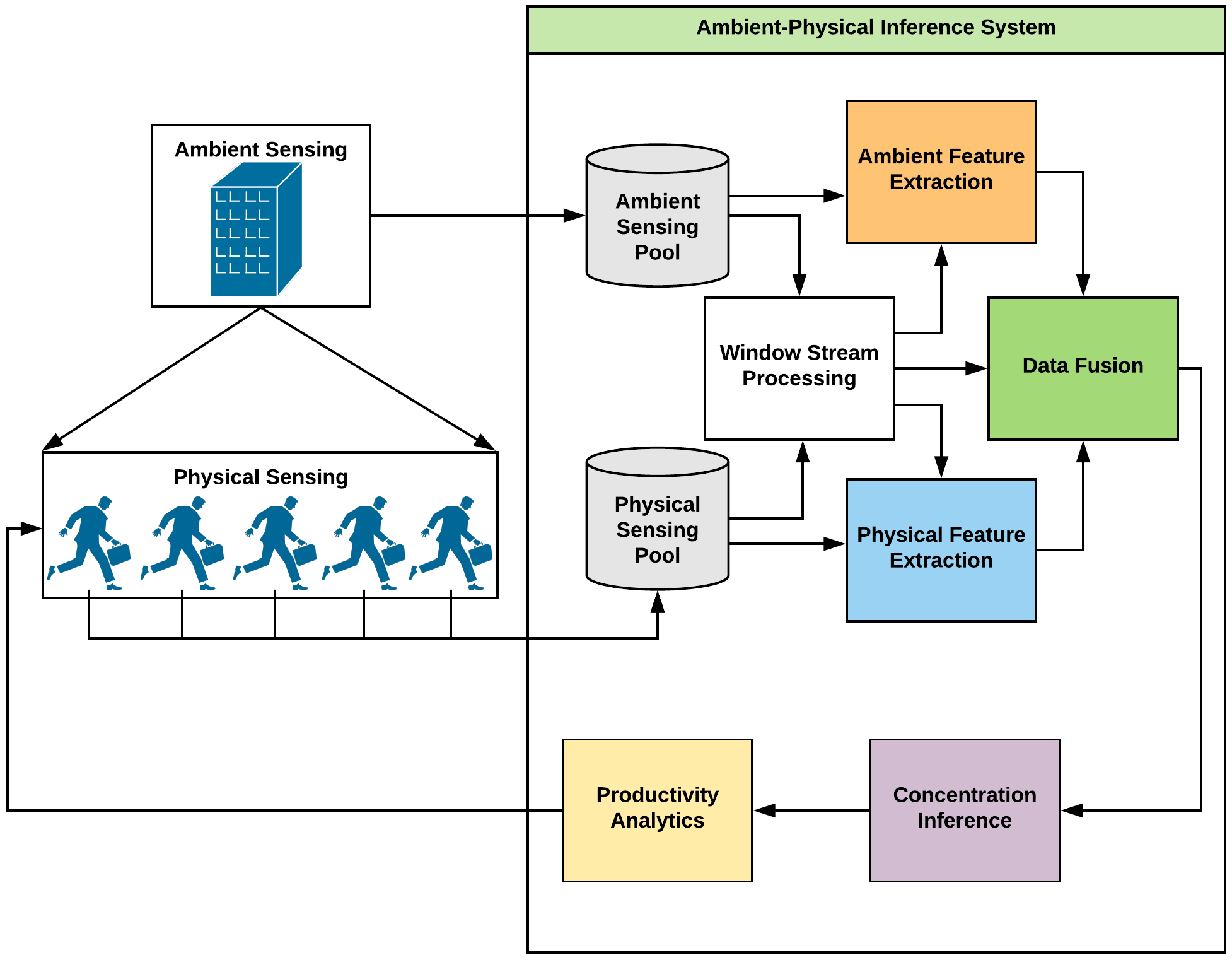}
  \caption{The Ambient-Physical system for concentration inference.}
  \label{fig:apisystem}
\end{figure}

In this section, we propose a system to infer the concentration of workers. Our system relates the concentration of workers in shared workplaces to physical activities and ambient environments while avoiding the privacy issues caused by other sensing technologies like cameras. The physical activities (i.e. movements) and ambient environment parameters are collected through passive sensing and processed by different modules of our system. As shown in Figure~\ref{fig:apisystem}, this system consists of the following key components: \emph{Sensing Pool}, \emph{Window Stream Processing}, \emph{Feature Extraction}, \emph{Data Fusion}, \emph{Concentration Inference} engine and \emph{Productivity Analytics} module.  

\subsection{Sensing Pool}
\label{sensingpool}
This sensing module is built on top of the knowledge from other research domains. Research in the behavioral psychology domain indicates a strong relationship between ambient environment parameters (e.g. temperature, air velocity, relative humidity, noise) and occupants' psychological performance in buildings \cite{rugg2010does, kamaruzzaman2011effect,banbury2005office, vimalanathan2014effect}. Research also has shown the contribution of body movements to the attribution of emotions \cite{zacharatos2014automatic}. Essentially, data from the ambient environment and workers' physical movement are collected and stored in the \emph{Sensing Pool} repositories for further data processing. The repositories of our ambient-physical inference system can be divided into two major sub-modules: ambient and physical sensing pools.

\textbf{Ambient Sensing Pool:} This module is responsible for collecting and storing data related to ambient environment parameters coming from the ambient sensors deployed in our pilot workplace infrastructure. The implementation of this pool was carried out in a Nectar cloud instance that integrates historical data from ambient sensors. In our experiments, we utilise Netatmo Weather Stations, which are equipped with sensors capable of measuring temperature, humidity, barometric pressure, noise level, and CO\textsubscript{2} density. The surrounding electromagnetic field caused by electronic devices were captured through participants' smartphone sensors. In this study, we installed 12 \textit{Netatmo} weather stations in each floor of the pilot sites. Netatmo sensors used in our experiment (i.e. Netatmo weather station) support a wireless connection range of 100 m (without obstacles). Since our testbed is approximately half of this range and we used multiple devices, our network configuration did not require any fingerprinting. The Netatmo sensors are Wi-Fi 802.11 b/g/n compatible (2.4GHz). To reproduce this work, one should refer to the latest specifications\footnote{\url{https://www.netatmo.com/en-eu/weather/weatherstation/specifications}}. 

\textbf{Physical Sensing Pool:} This module consists of a set of sensors to collect and store data that are sourced from participants' movements. Specifically, the physical sensing pool collects the concentration indicators of participants in terms of physical activities (e.g. non-regular movements, walking, sedentary behaviors) by leveraging their smartphone sensors including accelerometer, gyroscope, and pedometer. For seamless data collection, a smartphone application (both iOS and Android versions) called \textit{OpenSense} was developed by the research team. A frequency of 50 Hz was used to collect sensor streams. An index was built based on a unique identification of an individual for efficient data storage and retrieval. In our experiment, we utilised the smartphone's device ID together with a randomly generated participation ID to generate unique identifiers.

The data collected in sensing pools are processed and utilised by the \textit{feature extraction} module. Given a set of computed features and associated concentration labels, the \textit{concentration inference engine} module learns whether a participant is able to concentrate or not under certain circumstances for future inference. It should be noted that the passive Sensing Pool is not only limited to the ambient and physical sensors considered in this study but also extensible for future types of sensing that can be sourced from an office environment. A complete list of ambient-physical sensors are given in Table \ref{tab:sensors}. The statistical feature extraction from the data collected by the Sensing Pool is described in Section \ref{fe}.

\begin{table}[]
\centering
\caption{List of ambient-physical sensors.}
\label{tab:sensors}
\begin{tabular}{|l|p{6cm}|}
\hline
\textbf{Sensor types} & \textbf{Sensors}                                                                                                    \\ \hline

Ambient  & temperature, humidity, barometric pressure, noise level, magnetometer, CO\textsubscript{2} density \\ \hline
Physical  & accelerometer, gyroscope, pedometer  \\ \hline
\end{tabular}
\end{table}

\subsection{Window Stream Processing}
Based on the scanning of data in each sensing pool respectively, this operation will define the boundary of data that should be processed for the next phases: feature extractions and data fusion. Data from ambient and individual sensing will be treated in a sequential manner (indexed by timestamps). The window sizes considered in this study for ambient and physical feature extraction are 30 minutes (with 50\% overlaps) and 5 minutes respectively. Recent research presents application-specific techniques for optimal window size calculations \cite{liono2016optimal,sadri2017information}. Finding an optimal window size is not within the scope of this paper. 

\subsection{Feature Extraction}
\label{fe}
After defining the boundary for data points, the feature extraction process can be performed on the raw sensing data based on its respective rules. Essentially, the rules are based on the system designer in terms of what feature engineering techniques should be applied from the continuous streams of sensor data. In this case, we extract statistical features for all the ambient and physical factors described in Section \ref{sensingpool} to construct our ambient and physical feature sets (i.e. $F_a$ and $F_p$). These features are extracted from a temporal window (via sliding window model) on the sensing data. Specifically, we compute mean, median, standard deviation, maximum, minimum, inter-quartile range and root-mean-square of a factor for each temporal window.

\subsection{Data Fusion}
This component aligns the features extracted from sensing data (stored in ambient and physical sensing pools). Since each window has indexed timestamps, they can be aligned altogether and be used directly for concentration inference. 

\begin{figure}[h]
  \centering
  \includegraphics[width=0.8\columnwidth]{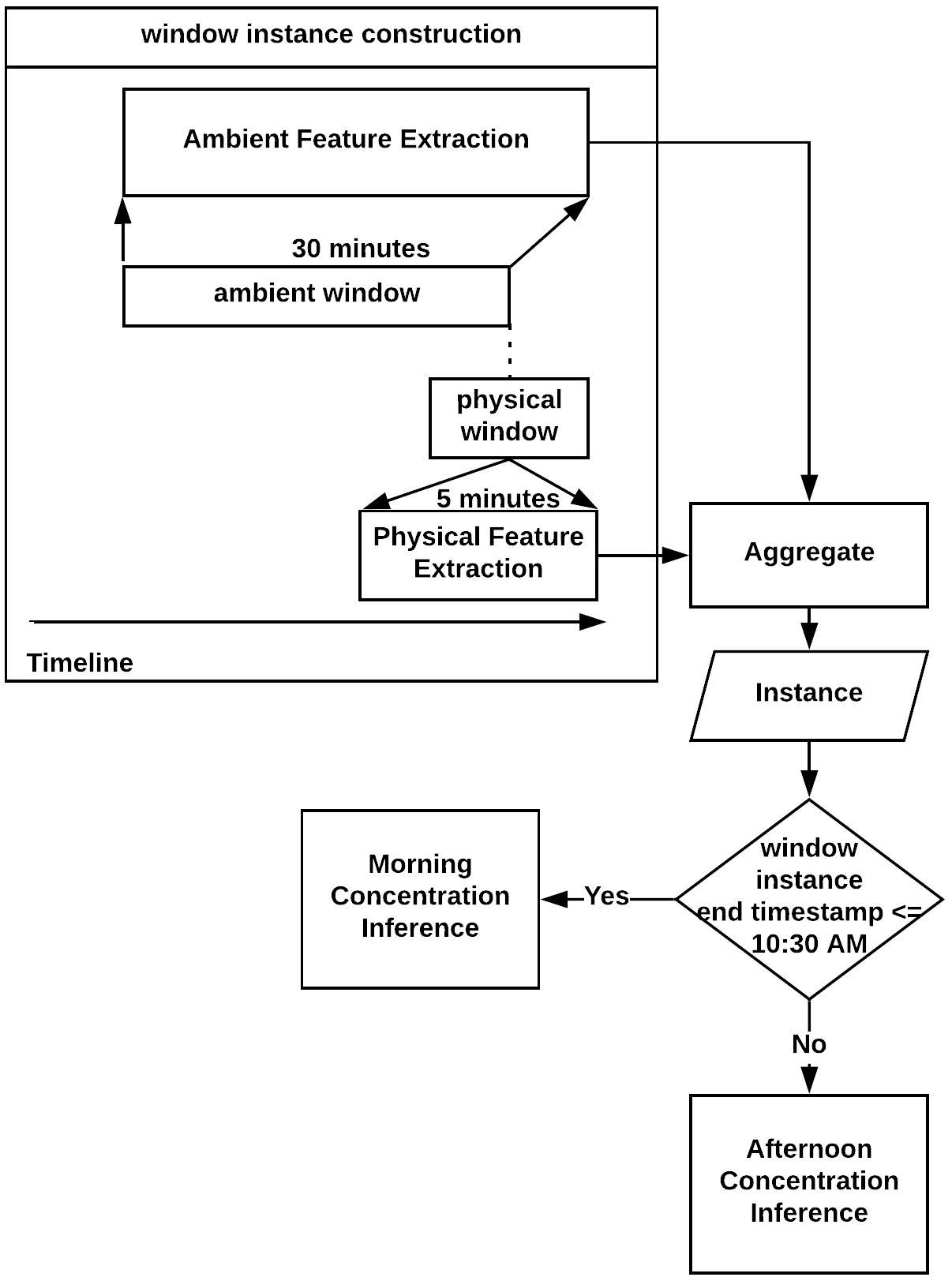}
  \caption{Data fusion for instance construction. }
  \label{fig:example-data-fusion}
\end{figure}

Each window used to extract physical sensing features is used as the base window for fusing the ambient sensing features. Hence, the window instance is constructed based on the exact alignment of upper bound of window boundary (i.e. end timestamp of its respective window). Consequently, the features extracted for ambient sensing data is then based on the last 30 minutes of the readings from all Netatmo sensor data on that open-office floor, where the end timestamp for ambient feature extraction lies in the middle of base window (i.e. the centroid of physical window, as shown in Figure~\ref{fig:example-data-fusion}). 

The construction of window instance is then aggregated based on the alignment of end timestamp of the base window. Consequently, this instance will be used for the inference purpose (whether for morning or afternoon concentration inference), depending on the predefined rules of ambient-physical inference system.  


\subsection{Concentration Inference Engine}
\begin{figure}[h!]
  \centering
  \includegraphics[width=0.5\columnwidth]{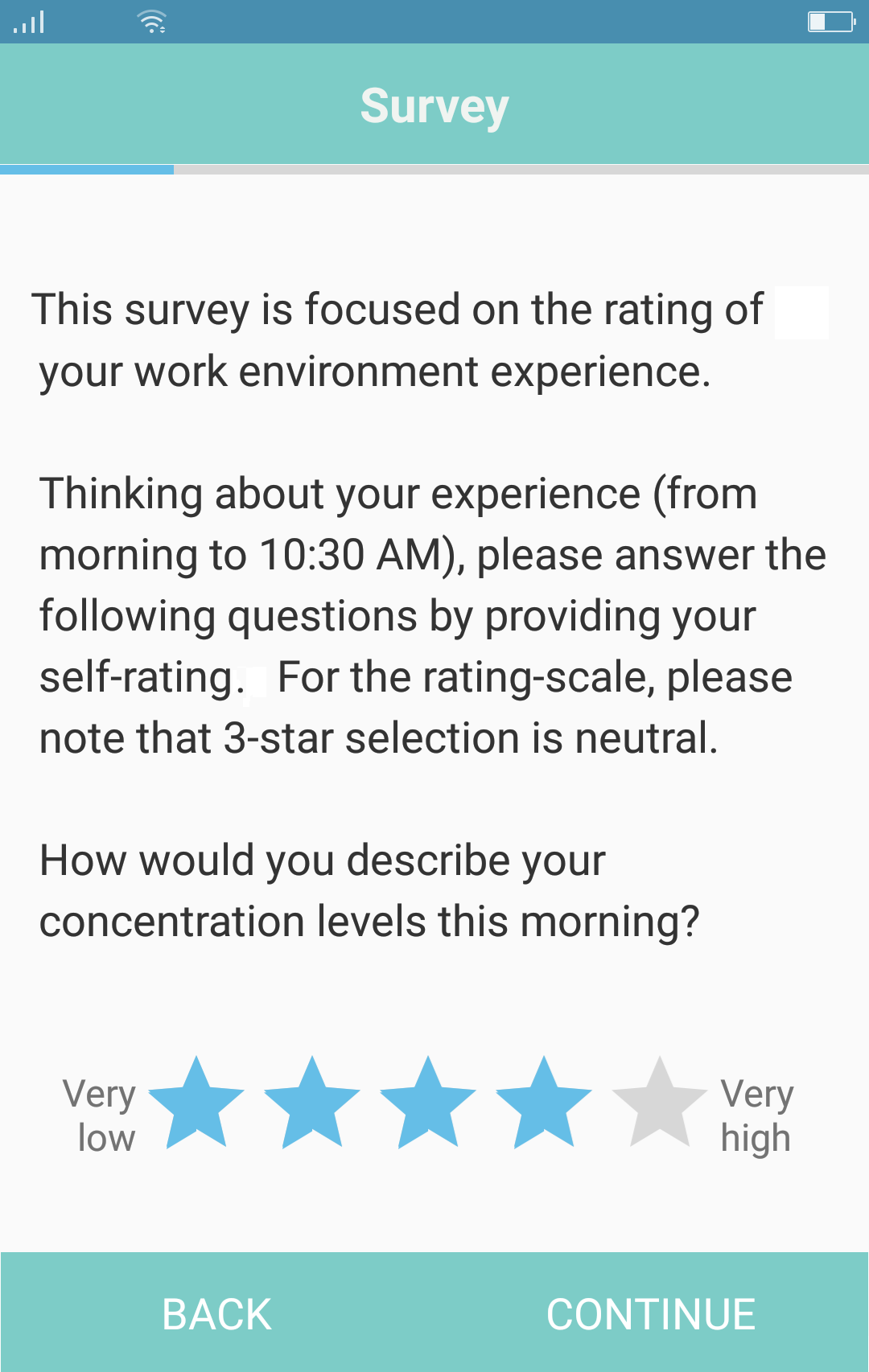}
  \caption{Concentration reporting through \textit{OpenSense}.}
  \label{fig:concentration-reporting}
\end{figure}

In this module, we validate the models that are built from the training data. Here, the self-reported ground truth was obtained through short (approx. 1 minute) surveys leveraging the in-app notifications of \textit{OpenSense} application. In our study, the participants reported their timestamped concentration levels ($\mathcal{L}_c$) twice a day (at 10:00 AM and 3:30 PM). The collection of ground truth concentration levels twice a day was decided based on the recommendation from a team of professional psychologists to keep the distractions caused by app notifications as minimal as possible. Figure \ref{fig:concentration-reporting} illustrates the concentration reporting screen of the \textit{OpenSense} application that was used by the participants during the study. Consequently, the unlabeled data that are pre-processed from previous steps will be used for inference purpose, to predict the concentration of workers. 

\subsection{Productivity Analytics Module}
This module will use the output of the concentration inference engine for further analytics purposes. This conceptual module is designed so that the proposed ambient-physical inference system is not only limited to concentration inference. Ideally, this module is a shared module that is integrated with other internal systems of an organisation. Although our study in this paper is focused on the prediction of worker concentration, we believe that it is only one of many indicators of worker productivity. Hence, there are tremendous challenges in this domain as the measurement of productivity is inherently difficult to address within current literature. 

The output of the productivity analytics module enables the operators (e.g. executives) to make informed decision that improves the overall productivity of the organization. Depending on the corporate policy and application design, the insights can be projected to these operators based on aggregation of teams, departments or branches. The workers could also be informed about the aggregated result of ambience and concentration map of the office. 

\section{Experiments and Evaluation}\label{sec-experiment}
\label{subsec:datacollection-and-labelling}

In this section, we discuss the experimental setup and implications of our developed ambient-physical system for concentration inference in the open-plan work environments.

\subsection{Open-plan Offices: Site-1 and Site-2}

\begin{figure}[h]
  \centering
  \includegraphics[width=0.9\columnwidth]{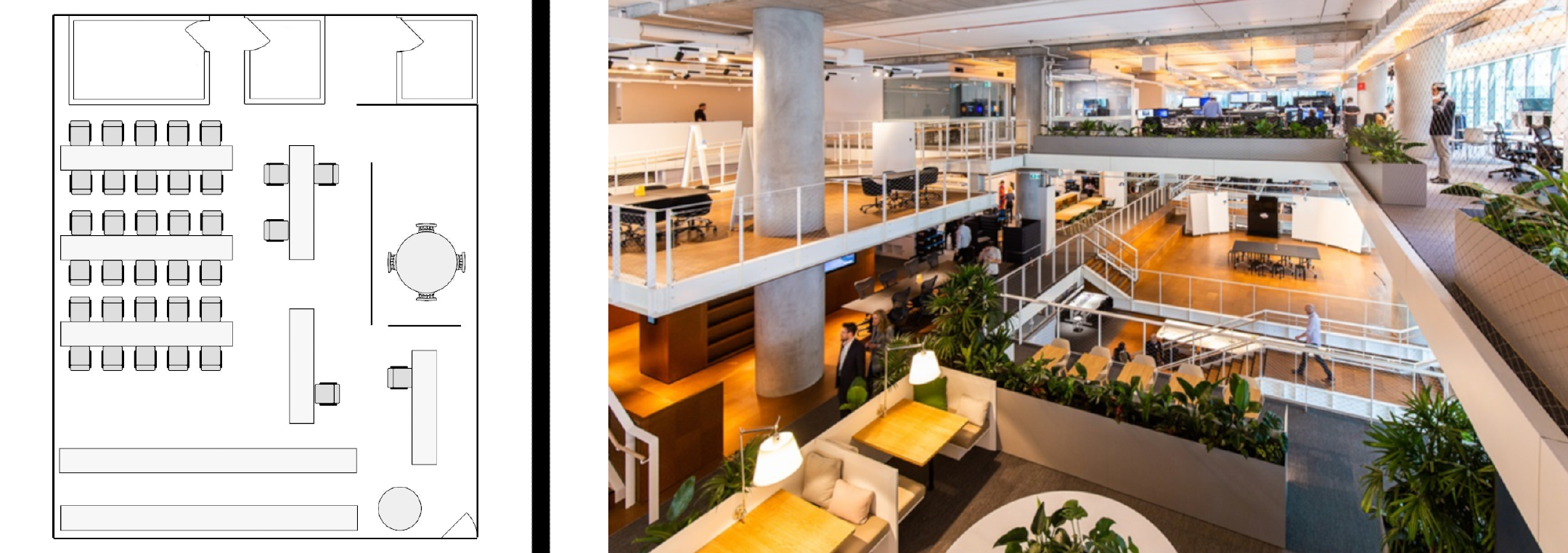}
  \caption{Illustrations of Site-1 (left) and Site-2 (right).}
  \label{fig:site1-site2-illustrations}
\end{figure}

To evaluate the applicability of our ambient-physical system on concentration inference, our experiment is systematically orchestrated based on the data collected over 4 weeks from two distinct office sites. In total, there were 31 willing volunteers participated in the data collection, consisting the sampling of 17 office workers in \textit{Site-1} and 14 office workers in \textit{Site-2}. Both offices practice open-plan work settings and differ in designs and layout. The \textit{Site-2} includes multiple hanging floors with a void connecting all of those while the \textit{Site-1} is an open-office with just a single floor (as illustrated in Figure~\ref{fig:site1-site2-illustrations}). The corporate participants in both sites work in the same organisation. In fact, they moved from \textit{Site-1} (old office) to \textit{Site-2} (new office). The data collection was performed within constrained time period (20$\approx$25 working weekdays) before they moved from \textit{Site-1} and once they settled in \textit{Site-2}. The ambient-physical sensor data collection is essentially comprised of the following tools: Netatmo Weather Stations (ambient-sensing) and OpenSense mobile app (ambient-physical sensing) installed on the personal smartphones of office workers. In such a scenario where employees moved from \textit{Site-1} to \textit{Site-2} due to the executive decision, it would influence the approaches in collecting ambient-physical sensor data, which was considered within our expectations. The first challenge in this moving-scenario is the interoperability issues of ambient sensing devices, which are caused by new infrastructure in \textit{Site-2} only supporting 5GHz wireless frequency devices. In fact, many of current sensing technologies still rely on the 2.4GHz wireless frequency to transfer the information through the network in real time. Therefore, the use of Netatmo Weather Stations was not prominent in \textit{Site-2}. Another challenge was the optimal placement of ambient-sensing devices that can cover the segregated and complex (multi-layered) spaces in \textit{Site-2} (assuming if we have built an independent mesh wireless network that supports 2.4GHz wireless frequency). In this case, we marked and labelled each zone (defined with ``zone-colour-encoding" mechanism, e.g. ``Violet" and ``Aqua"), and asked the participants for their sitting zone of the day (in morning app-survey during their occupancy in \textit{Site-2}). Note that each participant can practically sit anywhere in both sites, depending on their projects, activities and personal preference on working spaces. Consequently, we performed independent concentration inference experiments for both \textit{Site-1} and \textit{Site-2}. 

\subsection{Data Collection Protocol}
\label{dataCollection-preprocessing}

\begin{figure}[h]
  \centering
  \includegraphics[width=3.2in]{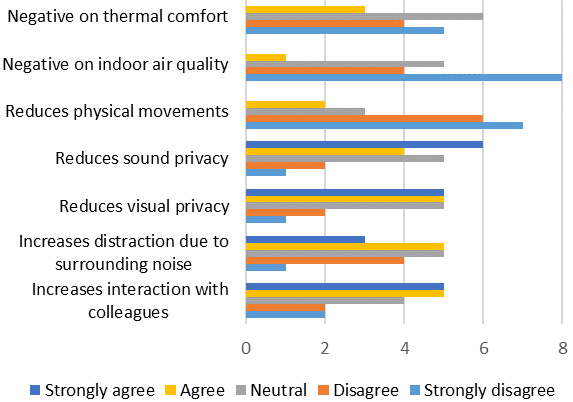}
  \caption{Perceptions on an open-plan workplace in comparison with a traditional workplace.}
  \label{fig:Perception-on-ABW}
\end{figure}

At the beginning of data collection, the anonymously recruited participants were asked to compare between different attributes of an open-plan office and that of a traditional workplace. Based on the initial responses, no significant concern was found regarding the negative impact on quality of indoor air in an open-plan office. This may be due to the use of a very high quality HVAC (Heating, Ventilation and Air Conditioning) system in our pilot sites. A small subset of the participants raised concern about reduction in physical movements (e.g. routine stretching). The majority of the participants highlighted the reduction in sound and visual privacy with increased noise distractions. However, as shown by several studies, participants agreed that an open-plan workplace environment increases their professional interaction with colleagues. Figure \ref{fig:Perception-on-ABW} illustrates the survey outcomes against different questions asked (in both \textit{Site-1} and \textit{Site-2}).

The participants were required to install the \textit{OpenSense} application and carry their mobile phones at all times while they are within the office hours (8:00 AM to 5:00 PM). This application senses smartphone sensor data in real-time from the participants' mobile devices which includes accelerometer, magnetometer, gyroscope, and pedometer. The participants also used \textit{OpenSense} application to report their timestamped concentration levels. It should be noted that the mobile sensor data were stored offline to provide participants with an option to review their data for increased privacy. The data were backed up automatically to our cloud storage every midnight. In the case of network failure during data backup, the data transmission pauses and restarts from where it was paused once the network becomes available. The smartphone models used by our participants during the data collection were released no earlier than the previous three years. During the course of data collection, no user complained about the battery drainage of their smartphones. Since geo-locations are not reported as part of our continuous data collection, the battery consumption should be reasonable for any latest release of smartphones.

\subsection{Experimental Setup}
\label{exp:setup}
To evaluate our ambient-physical concentration inference system, we carry out three sets of experiments as follows: 

\begin{itemize}
    \item \textit{Participant-independent concentration inference based on ambient features}. In this experiment, we used the ambient features ($F_a$) only for training and testing a set of classifiers for concentration inference.
    
    \item \textit{Participant-independent concentration inference based on physical features}. This experiment considered only the physical features ($F_p$) associated with our participants to investigate the concentration inference performance.
    
     \item \textit{Participant-independent concentration inference based on ambient and physical features together}. We computed an ambient-physical feature vector by concatenating ambient features with physical features. This combined feature vector was used to train and evaluate the performance of the concentration level ($\mathcal{L}_c$) inference.
\end{itemize}

\begin{figure}[t]
  \centering
  \includegraphics[width=0.95\columnwidth]{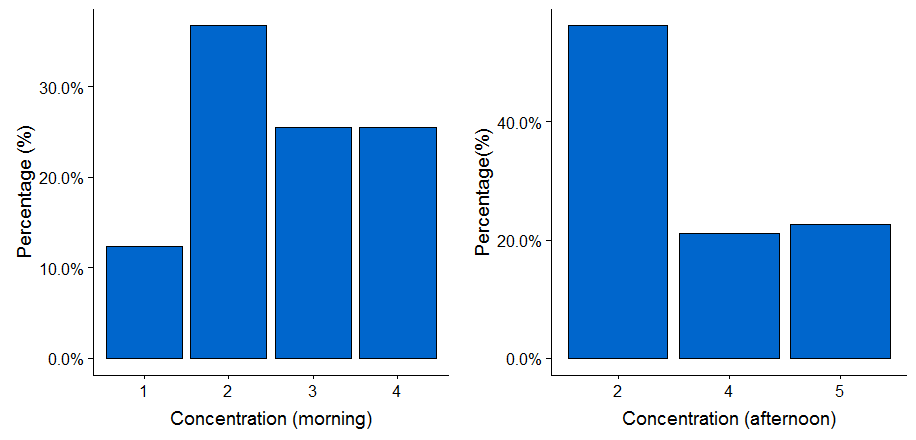}
  \caption{Concentration levels for an anonymous worker showing variations between morning and afternoon.}
  \label{fig:Concentration-distribution-E1L7}
\end{figure}

Note that the three sets of experiments as mentioned earlier in this section were conducted using four separate concentration datasets (i.e. two datasets per pilot site considering two different time slots of the day) denoted as \textit{Site-1: Morning}, \textit{Site-1: Afternoon}, \textit{Site-2: Morning} and \textit{Site-2: Afternoon}. The \textit{Morning} datasets contain morning-time  concentration of workers (i.e. between 8:00 AM and 10:30 AM) while the \textit{Afternoon} datasets contain afternoon time concentration levels (i.e. between 10:30 AM and 3:30 PM). The reason for doing so was inspired by the outcome of the initial survey which revealed that the concentration distribution varies significantly between morning and afternoon times. Figure \ref{fig:Concentration-distribution-E1L7} illustrates the distribution of concentration levels for an anonymous participant during morning and afternoon times of the data collection.

We consider concentration inference experiments using a single feature set (ambient or physical features) only in a participant-independent setting as two of our baselines. We compare the  concentration inference performance of our ambient-physical system using a combined feature set of ambient and physical features in a similar  participant-independent setting. 

In our experiments, we deployed a set of state-of-the-art classifiers including Gradient Boosting Trees, Decision Tree, Logistic Regression, Linear SVC, Stochastic Gradient Decent, XGboost, Random Forest, CatBoost, Na\"ive Bayes (NB), and $k$-Nearest Neighbor ($k$-NN). The scikit-learn~\cite{scikit-learn} machine learning library for Python was used to build these classifiers during our experimentation. 

In our Gradient Boosting Trees and Stochastic Gradient Decent implementations, we used the default parameters of scikit-learn. We used a tolerance parameter of 0.001 with a radial basis function (RBF) kernel for SVC classifier. A variant of Na\"ive Bayes algorithm called Gaussian Naive Bayes was used in our experiments. We used $k=10$ for our implementation of $k$-NN classifier. We considered n\_estimators=100 and max\_depth = 5 for Random Forest and XGBoost. Decision trees can be overfitted easily. In our decision tree implementations, we varied the depth of the trees between 2 and 10. We also restricted the minimum number of samples required to split an internal node by varying its value between 10 and 20. We reported the best results in our experiments. To handle the \textit{overfitting} in Catboost, the overfitting-detector type `Iter' was used along with 500 iterations. The loss function was `Multiclass' and `Accuracy' was chosen for custom loss metric. Note that finding the optimal parameter setting in our experiments was not within the scope of this research. 

\subsection{Evaluation}
\label{subsec:evaluation}

In our datasets, the reported concentration levels have an imbalanced distribution. Therefore, we adopted the stratified 10-fold cross-validation in our datasets. The stratification reduces the bias of over-represented concentration labels by ensuring approximately equal representation of each label across different folds of training and testing. In order to evaluate the performance of concentration inference using our ambient-physical system, we deploy accuracy as the measurement metric: 
 \begin{equation}
    {Accuracy}_{(\mathcal{L}_c)} = \frac{\textit{number of correct prediction}}{\textit{ total number of test cases}}
\end{equation}
where \textit{\# correct prediction} means the number of correctly predicted concentration labels, and \textit{\# total number of test cases} is the total number of test cases in the test set.

Tables ~\ref{table:morning-SITE-1-accuracy} and ~\ref{table:afternoon-SITE-1-accuracy} show the accuracies produced by a suite of eight classifiers when applied to the \textit{Site-1} datasets (i.e. \textit{Site-1: Morning} and \textit{Site-1: Afternoon}) while the performance matrices for \textit{Site-2} datasets (i.e. \textit{Site-2: Morning} and \textit{Site-2: Afternoon}) are summarized in Tables \ref{table:morning-SITE-2-accuracy} and \ref{table:afternoon-SITE-2-accuracy}. Note that `A', `P' and `A+P' represent the experiments with ambient features only, physical features only and combined ambient-physical features respectively.

\begin{table}[h!]
\centering
\caption{Concentration prediction accuracy (\%) by different classifiers in the \textit{Site-1: Morning} dataset using ambient (A), physical (P) and ambient-physical (A+P) features.}
\begin{tabular}{|l|l|l|l|}
\hline
\textbf{Model}             & \textbf{A}     & \textbf{P}     & \textbf{A + P} \\ \hline
Gradient Boosting Trees    & 50.83          & 48.51          & 51.16          \\ \hline
Decision Tree              & 52.01          & 44.62          & 49.11          \\ \hline
Logistic Regression        & 30.5           & 44.16          & 33.33          \\ \hline
Linear SVC                 & 27.33          & 34.32          & 25.54          \\ \hline
10-NN                     & 30.63          & 39.74          & 29.9           \\ \hline
Stochastic Gradient Decent & 28.25          & 29.24          & 33.8           \\ \hline
Naive Bayes                & 31.49          & 10.56          & 19.47          \\ \hline
Random Forest              & 53.05          & 45.16          & 54.27          \\ \hline
XGBoost                     & 79.33          & 60.97          & 87.35          \\ \hline
CatBoost                   & \textbf{91.97} & \textbf{71.56} & \textbf{97.16} \\ \hline
\end{tabular}
\label{table:morning-SITE-1-accuracy}
\end{table}

\begin{table}[h!]
\centering
\caption{Concentration prediction accuracy (\%) by different classifiers in the \textit{Site-1: Afternoon} dataset using ambient (A), physical (P) and ambient-physical (A+P) features.}
\begin{tabular}{|l|l|l|l|}
\hline
\textbf{Model}             & \textbf{A}   & \textbf{P}   & \textbf{A + P} \\ \hline
Gradient Boosting Trees    & 52.89          & 40.26          & 55.12            \\ \hline
Decision Tree              & 52.63          & 38.26          & 51.87            \\ \hline
Logistic Regression        & 27.82          & 36.7           & 27.3             \\ \hline
Linear SVC                 & 25.94          & 35.9           & 29.89            \\ \hline
10-NN                        & 20.24          & 33.1           & 20.38            \\ \hline
Stochastic Gradient Decent & 33.8           & 26.48          & 28.27            \\ \hline
Naive Bayes                & 27.68          & 16.91          & 19.01            \\ \hline
Random Forest              & 54.28          & 40.62          & 57.81          \\ \hline
XGBoost                     & 77.63          & 59.36          & 85.77          \\ \hline
CatBoost                   & \textbf{85.81} & \textbf{66.88} & \textbf{96.56}   \\ \hline
\end{tabular}
\label{table:afternoon-SITE-1-accuracy}
\end{table}

We can see from Tables \ref{table:morning-SITE-1-accuracy} - \ref{table:afternoon-SITE-2-accuracy} that the Catboost classifier outperforms other classifiers significantly. It should be noted that any gradient boosting algorithm could be a good fit for problems like ours where we don't have the luxury of big sized data and we don't want to overfit a model for better accuracy. Moreover, CatBoost is faster since it implements a symmetric tree structure. It also divides a given dataset into random permutations before applying boosting which can stop overfitting. There are dedicated overfitting detectors in Catboost classifier. We used overfitting-detector type ‘Iter’ in our experiment. We also can see from Tables \ref{table:morning-SITE-1-accuracy} - \ref{table:afternoon-SITE-2-accuracy} that the combination of ambient and physical features produces substantial improvement in accuracy scores compared to ambient features only or physical features only in a participant-independent scenario. For instance, the improvements are between 5 - 10\% over ambient features and 26 - 30\% over physical features in both \textit{Site-1} datasets. The experiment with Catboost using only ambient or  physical features produce similar prediction accuracies in \textit{Site-2} datasets, however, the combined feature set (i.e. A + P) improves the prediction accuracies by 10 to 12\%.

\begin{table}[h]
\centering
\caption{Concentration prediction accuracy (\%) by different classifiers in the \textit{Site-2: Morning} dataset using ambient (A), physical (P) and ambient-physical (A+P) features.}
\begin{tabular}{|l|l|l|l|}
\hline
\textbf{Model}             & \textbf{A}     & \textbf{P}     & \textbf{A + P} \\ \hline
Gradient Boosting Trees    & 38.32          & 44.97          & 43.02          \\ \hline
Decision Tree              & 39.22          & 43.77          & 42.97          \\ \hline
Logistic Regression        & 44.67          & 44.87          & 44.47          \\ \hline
Linear SVC                 & 44.77          & 43.72          & 42.17          \\ \hline
10-KNN                     & 38.92          & 40.67          & 41.82          \\ \hline
Stochastic Gradient Decent & 37.67          & 30.42          & 33.57          \\ \hline
Naive Bayes                & 32.57          & 10.11          & 15.51          \\ \hline
Random Forest              & 40.15          & 44.27          & 43.82          \\ \hline
XGBoost                     & 60.13          & 58.29          & 69.84          \\ \hline
CatBoost                   & \textbf{66.43} & \textbf{65.53} & \textbf{75.99} \\ \hline
\end{tabular}
\label{table:morning-SITE-2-accuracy}
\end{table}

\begin{table}[h]
\centering
\caption{Concentration prediction accuracy (\%) by different classifiers in the \textit{Site-2: Afternoon} dataset using ambient (A), physical (P) and ambient-physical (A+P) features.}
\begin{tabular}{|l|l|l|l|}
\hline
\textbf{Model}             & \textbf{A}     & \textbf{P}     & \textbf{A + P} \\ \hline
Gradient Boosting Trees    & 46.92          & 44.02          & 47.52          \\ \hline
Decision Tree              & 47.91          & 44.73          & 47.49          \\ \hline
Logistic Regression        & 44.24          & 43.72          & 43.91          \\ \hline
Linear SVC                 & 45.01          & 41.91          & 43.13          \\ \hline
10-KNN                     & 45.93          & 44.59          & 47.73          \\ \hline
Stochastic Gradient Decent & 42.33          & 41             & 41.17          \\ \hline
Naive Bayes                & 31.82          & 9.13           & 9.55           \\ \hline
Random Forest              & 47.98          & 44.86          & 48.20          \\ \hline
XGBoost                     & 60.39          & 57.72          & 70.55          \\ \hline
CatBoost                   & \textbf{66.16} & \textbf{64.31} & \textbf{76.07} \\ \hline
\end{tabular}
\label{table:afternoon-SITE-2-accuracy}
\end{table}

\begin{figure}[h]
  \centering
  \includegraphics[width=\columnwidth]{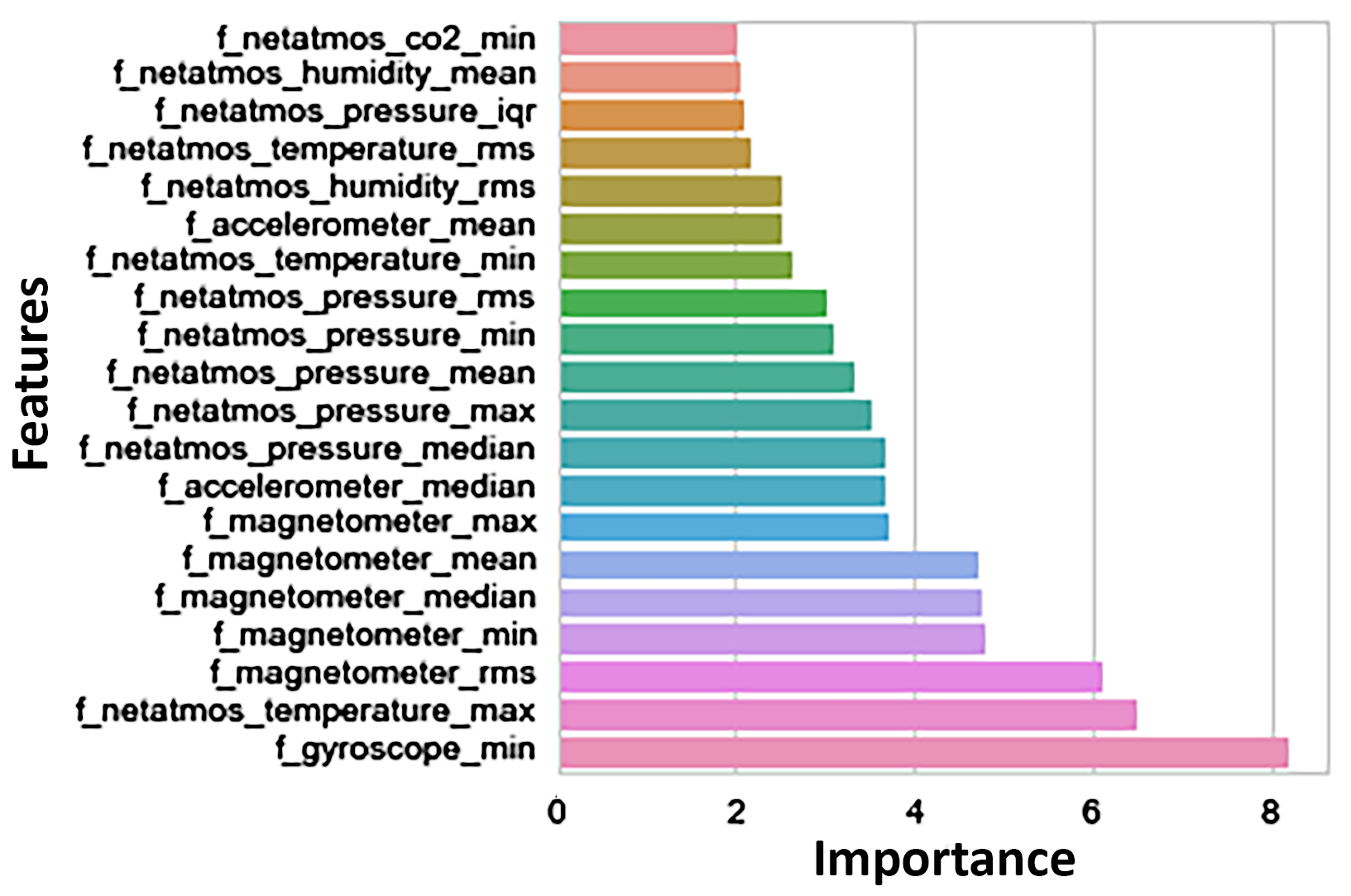}
  \caption{Important ambient-physical features by Catboost.}
  \label{fig:top-features}
\end{figure}

Further, we explore several ambient-physical indicators which were considered to be useful for concentration inference. The top-20 important ambient and physical features considered by a Catboost classifier to build a model is illustrated in Figure~\ref{fig:top-features}. This includes features computed from magnetometer, accelerometer, gyroscope, humidity, temparature, CO\textsubscript{2} and indoor air pressure. We are intrigued by the consensus of our participants highlighting that their perceived concentration level is not impacted by the ambient temperature and air quality in the office environments. However, these two factors seem to have impact on perceived concentration levels during the course of data collection in both sites.

\subsection{Discussion and Implications}\label{sec-discussion}

\begin{figure}[h!]
  \centering
  \includegraphics[width=\columnwidth]{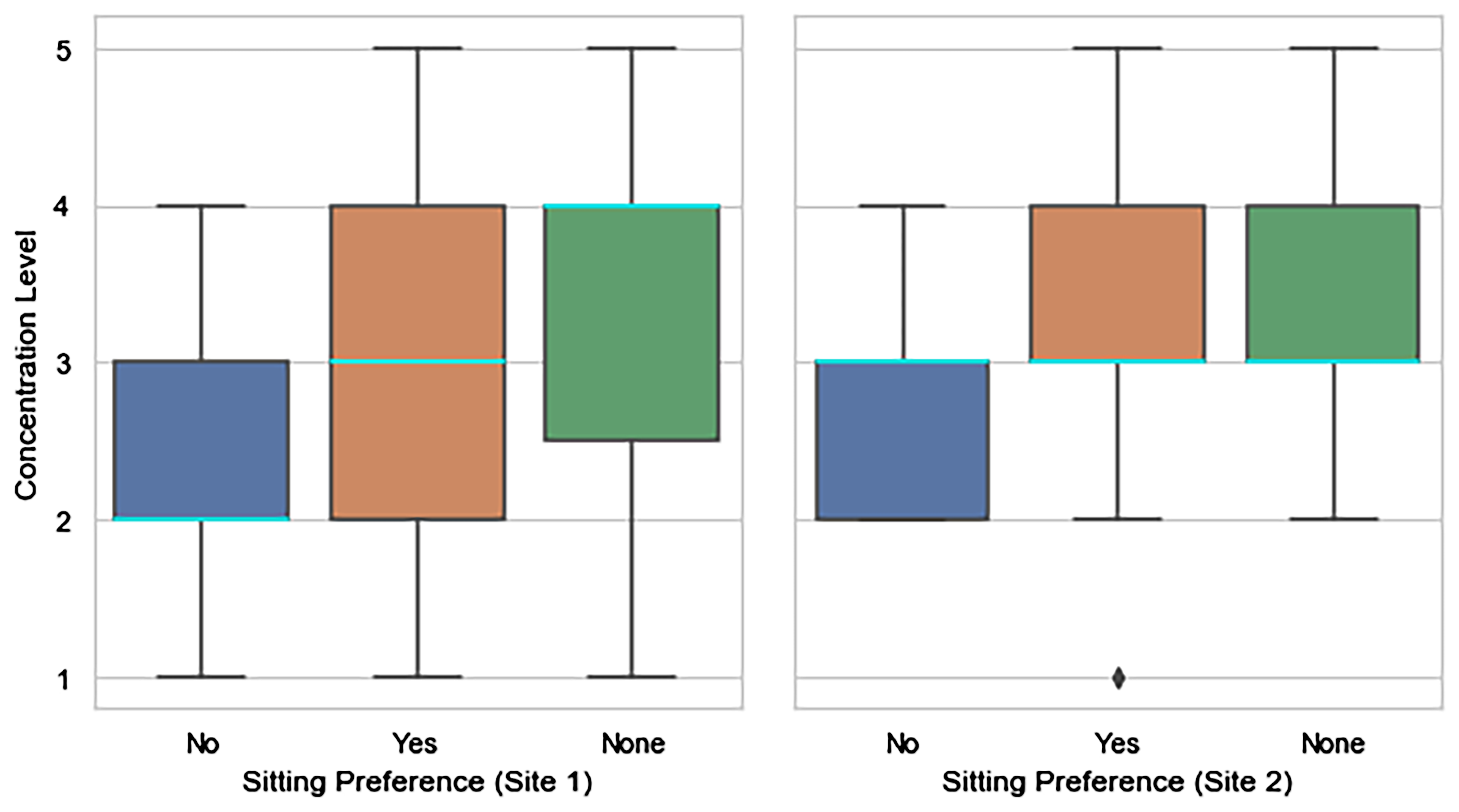}
  \caption{Sitting preference vs concentration level.}
  \label{fig:comparison-concentration-sittingpreference}
\end{figure}

As stated in the previous section, the collected data exhibits higher prediction performance for \textit{Site-1} in compared to \textit{Site-2}. This could be due to the difference in size, design and layout between our pilot sites. We also investigate the effect of different external factors associated with perceived concentration of workers. Figure \ref{fig:comparison-concentration-sittingpreference} shows that workers' sitting preference could influence their perceived concentration level. Although survey data in \textit{Site-2} implies a subtle change of concentration distribution from \textit{Site-1}, having the participants to sit in their preferred zones is extremely important for their well-being and overall productivity. This subtle change can also be caused by more available selections of spaces to work in \textit{Site-2}. 

\begin{figure}[t]
  \centering
  \includegraphics[width=\columnwidth]{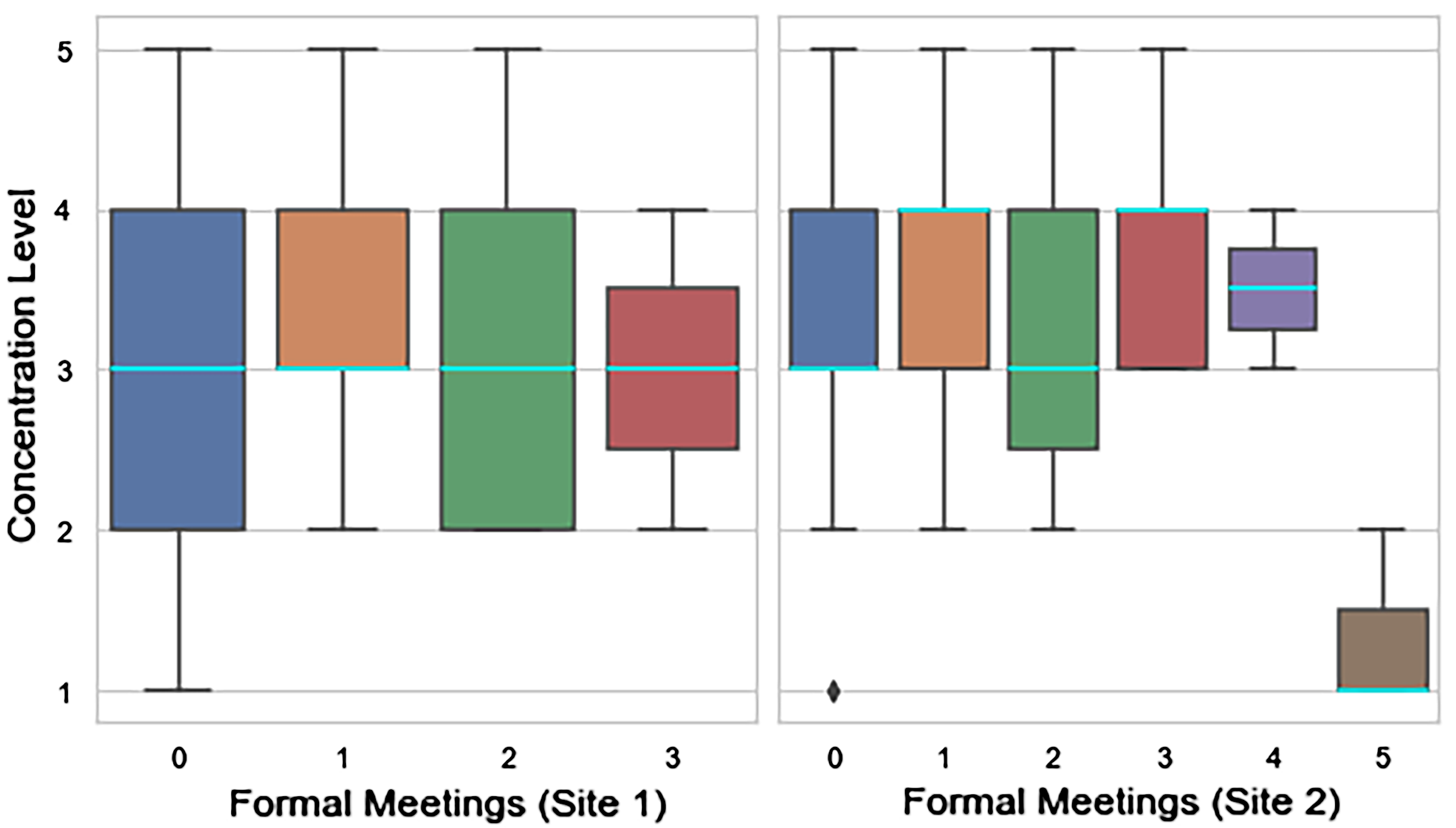}
  \caption{Number of formal meetings vs concentration level.}
  \label{fig:comparison-concentration-formalmeetings}
\end{figure}

\begin{figure}[t]
  \centering
  \includegraphics[width=\columnwidth]{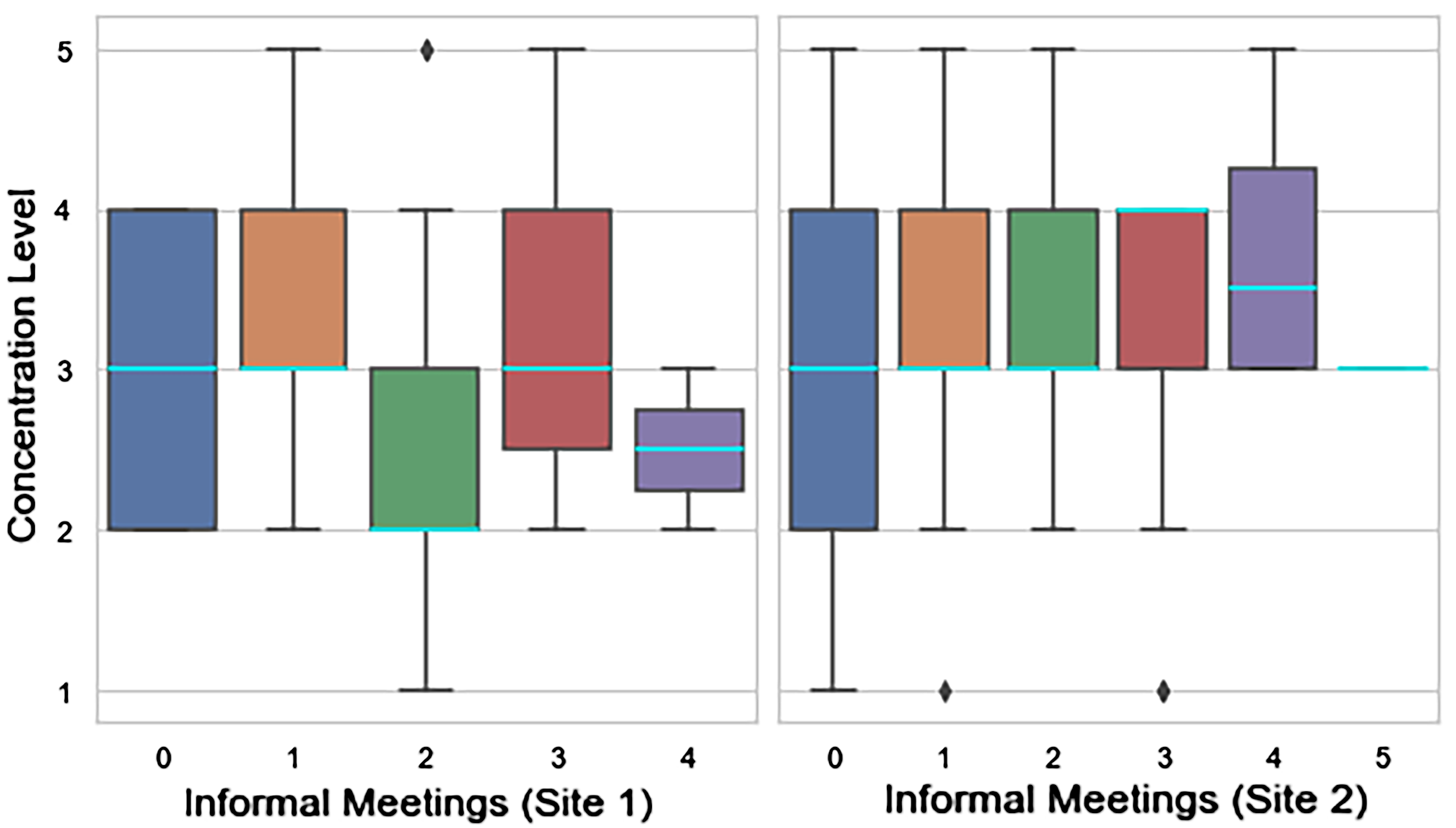}
  \caption{Number of informal meetings vs concentration level.}
  \label{fig:comparison-concentration-informalmeetings}
\end{figure}

The numbers of formal and informal meetings can also influence the concentration level. As can be seen from Figure \ref{fig:comparison-concentration-formalmeetings}, the formal meetings in \textit{Site-1} seem to exhibit a median of 3 (neutral) concentration level for different number of meetings between 0 and 3. Although the workspaces in \textit{Site-2} were designed and aimed to increase workers' productivity, the number of formal meetings seems to have varying impact on median concentration level. The concentration reaches it's lowest with 5 formal meetings. However, the increased number of informal meetings in \textit{Site-2} seems to exhibit a slightly better median concentration level compared to \textit{Site-1} (see Figure~\ref{fig:comparison-concentration-informalmeetings}).

\begin{figure}[h!]
  \centering
  \includegraphics[width=\columnwidth]{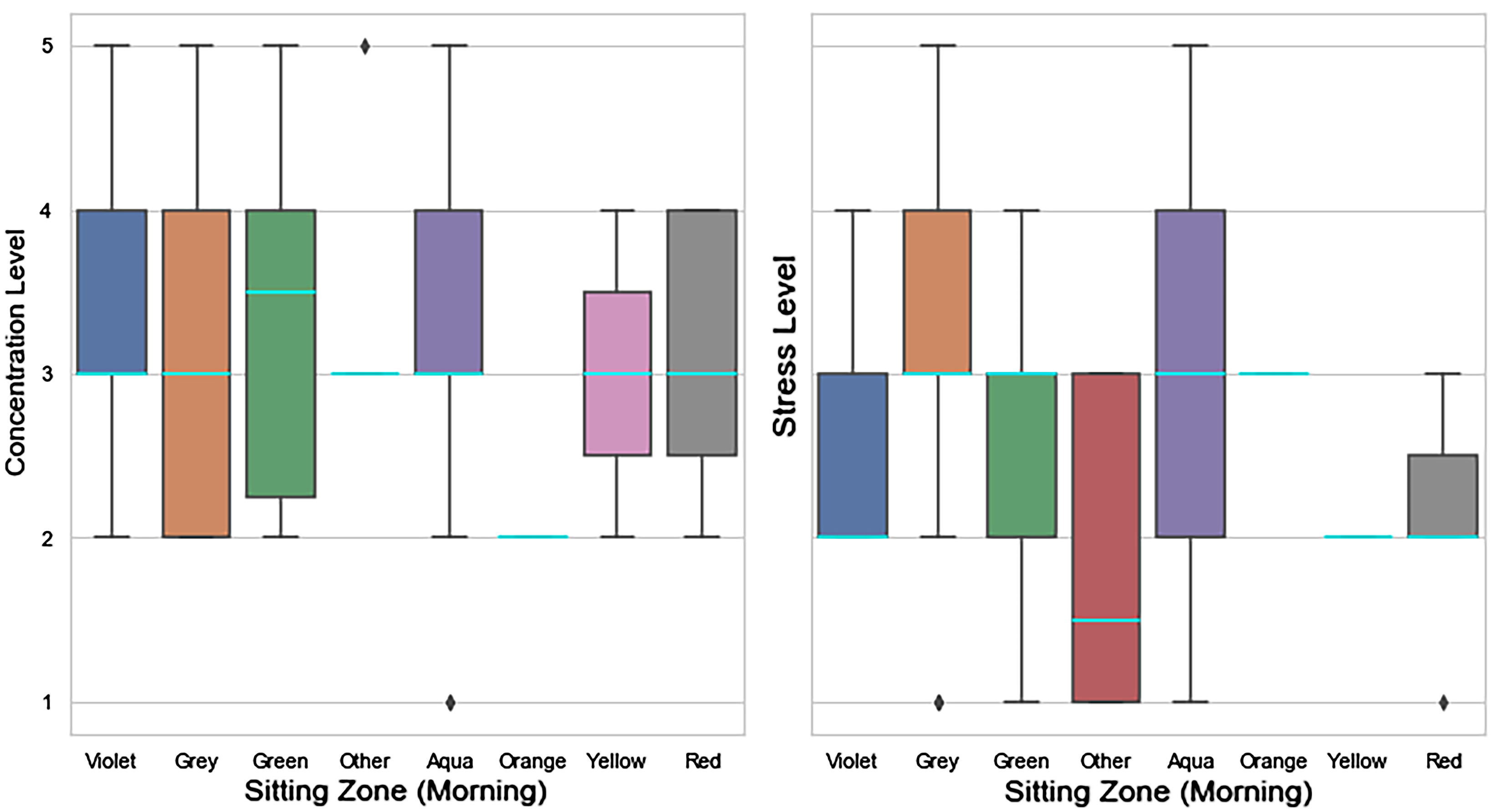}
  \caption{Concentration and stress level in different sitting zones.}
  \label{fig:concentration-stress-sittingzone-site2}
\end{figure}

The sitting zones are small areas within the open-plan office floor. We collected sitting zone information of participants in the \textit{Site-2}. We identify that the perceived median concentration and stress may vary between morning and afternoon even when participants sit in similar location. Figure~\ref{fig:concentration-stress-sittingzone-site2} shows the concentration and stress in different color-coded zones in \textit{Site-2}. This information can be utilised to optimise the office layout for increased productivity. 


\begin{figure*}[t]
  \centering
  \includegraphics[width=0.85\textwidth]{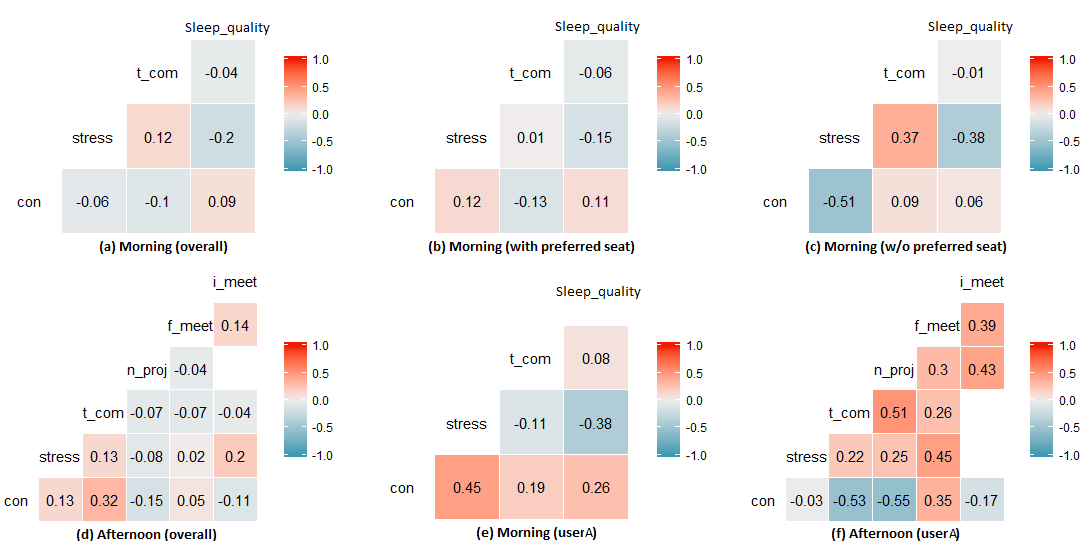}
  \caption{Correlation Matrix showing Pearson's correlation among different factors: (a) Overall Morning data, (b) Morning data when participants got preferred seat, (c) Morning data when participants did not get preferred seat, (d) Overall Afternoon data, (e) Overall Morning data (anonymous user A), (f) Overall Afternoon data (anonymous user A).}
  \label{fig:cap-1-afternoon-survey-corrMatrix}
\end{figure*}

We found that some participant workers exhibit different ambient-physical patterns which make the inference task challenging. This may be due to the absence of some signals in our ambient-physical system (e.g. signals that represents rare concentration traits). However, our system is not rigid towards the selected signals only and can be extended when relevant features become available. In this study, we only used passive sensing to capture different ambient-physical signals.

We also identified that additional inputs such as perceived stress, thermal comfort, sleep quality, number of formal meetings, informal meetings and active project involvements can be leveraged to conduct further analysis. We collected these additional factors through a one-off survey at the start of the day. It should be noted that the collection of these inputs requires an extensive amount of \emph{active} sensing (i.e. active participation of workers). Moreover, our developed system can achieve high accuracy in concentration inference utilizing a combination of \emph{passive} ambient-physical sensors. Hence, these factors are not considered to build off our prediction module. Next, we conduct an overall Pearson's correlation analysis among perceived concentration and all of these external factors. Figure \ref{fig:cap-1-afternoon-survey-corrMatrix} (a) shows a participant-independent correlation matrix devised considering the morning times where no significant correlation is found among concentration (con), stress, thermal comfort (t\_com) and sleep quality. We also analyzed if there is any association among these features considering their seating arrangements (e.g. preferred seat such as close to window, workstation with two screens, close to team members). As can be seen from Figure \ref{fig:cap-1-afternoon-survey-corrMatrix} (b)-(c), no association among these variables was found when the participants had a preferred seat. In the case of a seat which is not preferred by the participant, there was a positive correlation with their perceived stress which eventually has a strong negative correlation with their perceived concentration levels. We also found that a good quality sleep is very important to reduce stress at work. The analysis also shows that thermal comfort is positively correlated with overall concentration in the afternoon while no correlation was found between concentration, number of active project participation (n\_proj), number of formal meeting (f\_meet) and number of informal meeting (i\_meet) as can be seen from Figure \ref{fig:cap-1-afternoon-survey-corrMatrix} (d).

\begin{figure}[h!]
  \centering
  \includegraphics[width=\columnwidth]{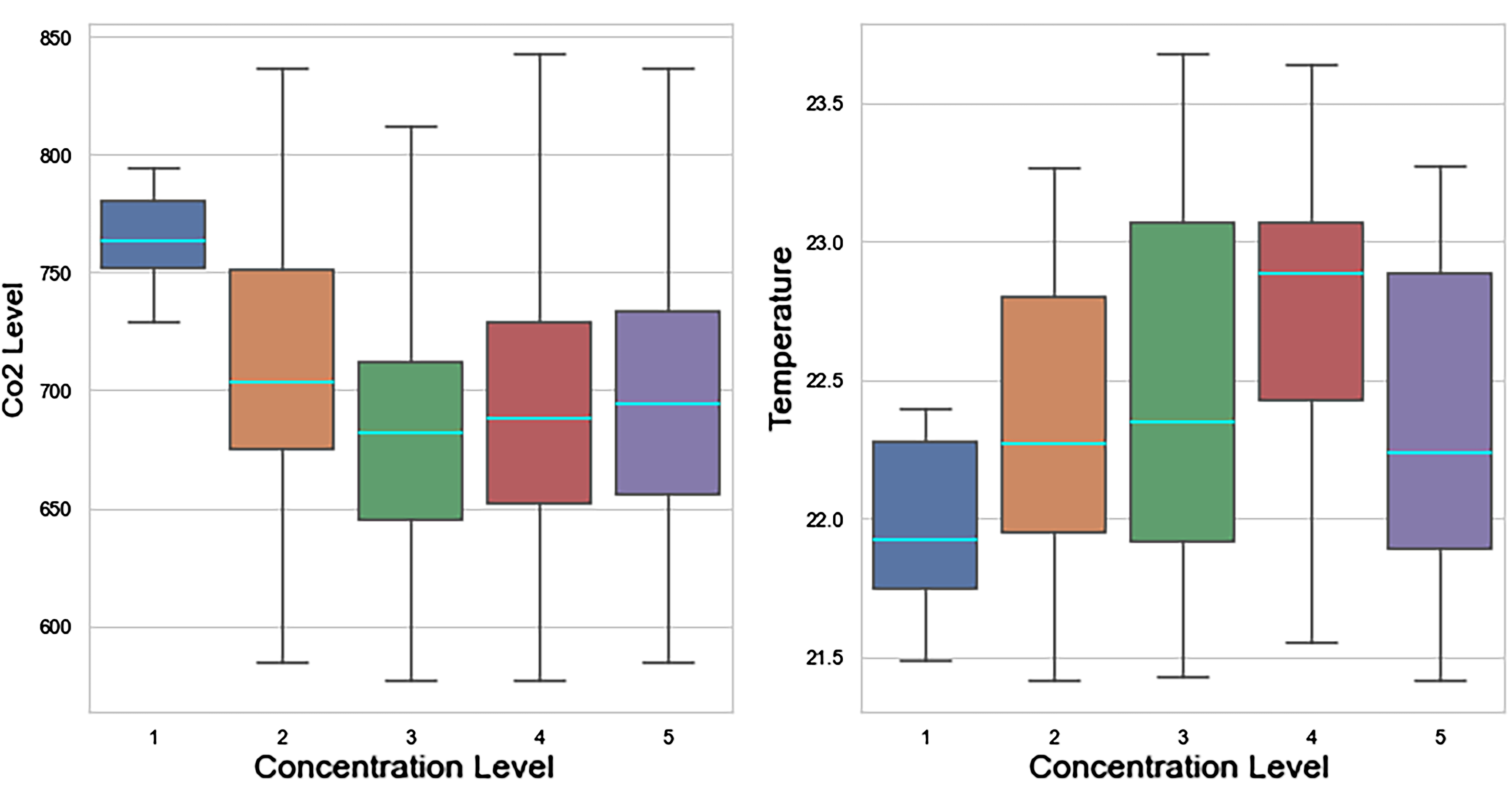}
  \caption{Ambient sensors: CO\textsubscript{2} density and temperature against concentration level for \textit{Site-1}.}
  \label{fig:concentration-co2-temperature-site1}
\end{figure}

To conduct further analysis, we found that the morning-time stress of some participant workers (e.g. user A) is influenced by the quality of sleep in the previous night as can be seen from the correlation matrix in Figure \ref{fig:cap-1-afternoon-survey-corrMatrix} (e). An investigation with the afternoon-time correlation matrix of some participants (e.g. user A) pointed to a strong negative correlation between concentration and number of active project participation which leads to a decline in thermal comfort as can be seen from Figure \ref{fig:cap-1-afternoon-survey-corrMatrix} (f). We identified that the higher number of formal meetings increases the stress level of some particular participants. We also found that both higher CO\textsubscript{2} density and lower temperature (i.e. 21.5-22.5 $^{\circ}$C) can lead to lower concentration level for some occupants (Figure  \ref{fig:concentration-co2-temperature-site1}). 

The findings in this paper provide the semblance of a mechanical relationship between perceived concentration and ambient-physical aspects of a workplace that in many cases can be controlled. For managers and executives, such a relationship can provide a framework through which to allocate resources both physically in an office and financially in terms of facilities and overheads. From a more general built environment perspective, such a relationship allows for deeper understanding of how humans are influenced by the world around them. Such an understanding leads to the potential optimisation of the modern office and private work environments. A major limitation to this however is the variability of modern offices, companies and workers. Data was collected from two sites of an engineering consultancy firm. Varying such factors may have high influence on the correlations found in the data. 

The study presented in this paper has several implications in the domain of open-plan workplace and beyond. Some key implications are summarised as follows:
\begin{itemize}
    \item Perceived concentration inference of workers can enable informed decision making regarding workplace design and layout, for instance the placement of certain teams in certain locations in the open-plan workplace.
    \item  Concentration inference can help identify factors associated with particular concentration traits in the workplace.
    \item It can help employees with rare concentration disorder condition while they are at work.
    \item Providing on-demand personalized solution to increase concentration level which will eventually increase the overall productivity of the organization. 
    \item Provide management with a clearer understanding of the ambient and physical factors that influence perceived concentration and certain behaviors in the workplace.
    \item Enhance managements' real-time understanding of worker satisfaction and productivity.
\end{itemize}

\section{Conclusions and Future Work}\label{sec-conclusion}

Concentration management in an aggregated manner can provide valuable insight while designing and managing an open-plan office. Concentration inference is important here, and has many applications, ranging from informed decision making to increased overall organizational productivity to work-zone recommendation for employees with rare medical conditions.  We presented an ambient-physical system to infer concentration of open-plan office workers. The core of our ambient-physical system is a pervasive sensing module which utilizes a set of ambient-physical sensors. This module enables ambient-physical feature computation for concentration inference in an open-plan work environment through sensory data fusion.

We also extracted exceptional patterns from a small number of participants where the concentration traits are influenced by additional factors (e.g. sitting location, number of active project involment, and number of meetings). To analyse such participants, we collected data through separate survey. Future research may address the challenge of collecting these data through the development of appropriate passive sensors. Also the exploration of transfer learning techniques can be considered in the future research. Future research also could investigate this problem as a sequential prediction task. Since it will require ground-truth concentration labels in a time-series manner, a non-obtrusive data collection protocol needs to be designed which will not compromise the concentration of workers during data collection.

\ifCLASSOPTIONcompsoc
  \section*{Acknowledgments}
\else
  \section*{Acknowledgment}
\fi

This research was supported by Arup and RMIT Enabling Capability Platform through the provision of an 'Opportunity Funding' scheme (no. 17073).

\ifCLASSOPTIONcaptionsoff
  \newpage
\fi

\bibliographystyle{IEEEtran}
\bibliography{Bibliography.bib}

%

\begin{IEEEbiography}[{\includegraphics[width=1in,height=1.25in,clip,keepaspectratio]{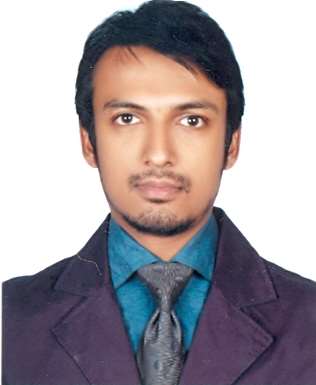}}]{Mohammad Saiedur Rahaman}
received his Ph.D. degree from School of Science (Computer Science \& IT), RMIT University, Melbourne, VIC, Australia. Currently, he is with the same school at RMIT University as a Research Fellow. Dr. Rahaman served as a co-chair of IEEE PerAwareCity 2020 and TPC member of PerCom Industry Track 2020. His current research interests include IoT, machine learning, spatio-temporal pattern mining, and context recognition in pervasive environments.
\end{IEEEbiography}

\begin{IEEEbiography}[{\includegraphics[width=1in,height=1.25in,clip,keepaspectratio]{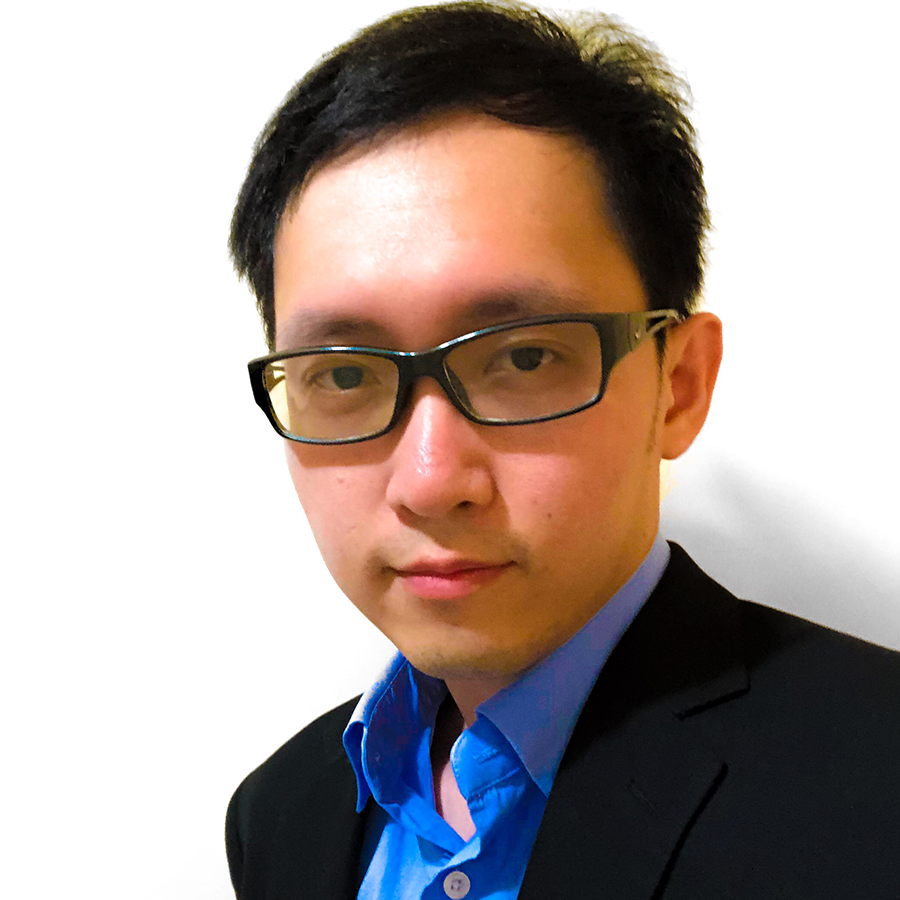}}]{Jonathan Liono} received his Ph.D. degree from School of Science (Computer Science \& IT), RMIT University, Melbourne, VIC, Australia. He is currently a Senior Data Scientist at AiDA Technologies, Singapore. He has served as Publication Co-Chair of MobiCase 2015. Recently, he has also served as Publicity Chair of the 2020 IEEE PerCom. His research interests include ubiquitous data mining, mobile context and activity recognition, spatio-temporal pattern mining and dynamic mobility pattern.  
\end{IEEEbiography}


\begin{IEEEbiography}[{\includegraphics[width=1in,height=1.25in,clip,keepaspectratio]{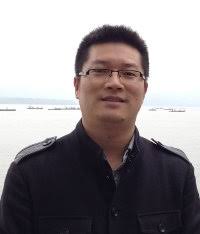}}]{Yongli Ren}
received the PhD degree in Information Technology from Deakin University, Australia. He is currently a Senior Lecturer of computer science and IT with the School of Science, RMIT University, Australia. His research interests include recommender systems, log analysis, user profiling, and data mining. He received the Alfred Deakin Medal for Doctoral Thesis in 2013 from Deakin University and the Best Paper Award from the IEEE/ACM ASONAM 2012 Conference.
\end{IEEEbiography}

\begin{IEEEbiography}[{\includegraphics[width=1in,height=1.25in,clip,keepaspectratio]{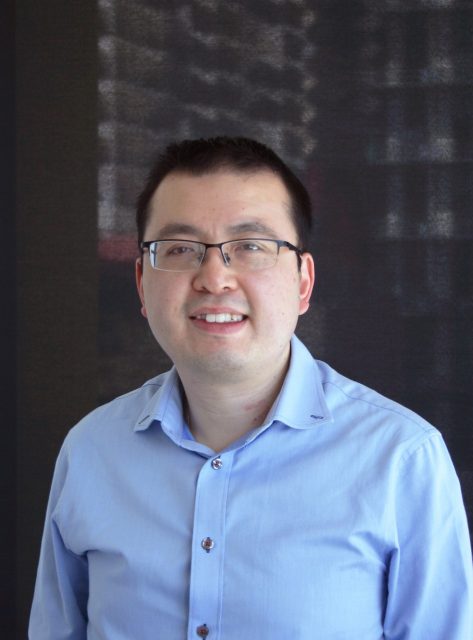}}]{Jeffrey Chan} is a senior lecturer at RMIT University.  He completed his BEng/BSci (Hons) and PhD at the University of Melbourne, Australia and was a senior postdoctoral research at the Digital Enterprise Research Institute in Galway, Ireland.  He has published more than 90 publications in machine learning, social network analysis, recommendation and data driven optimisation, in venues such as TPAMI, TKDE, DMKD, KDD, ICDM, AAAI and IJCAI.  He has served on various conference organising committees, such as IJCAI and ASONAM.  He has also won best paper award in the ACM International Conference on Web Science  in 2011.
\end{IEEEbiography}

\begin{IEEEbiography}[{\includegraphics[width=1in,height=1.25in,clip,keepaspectratio]{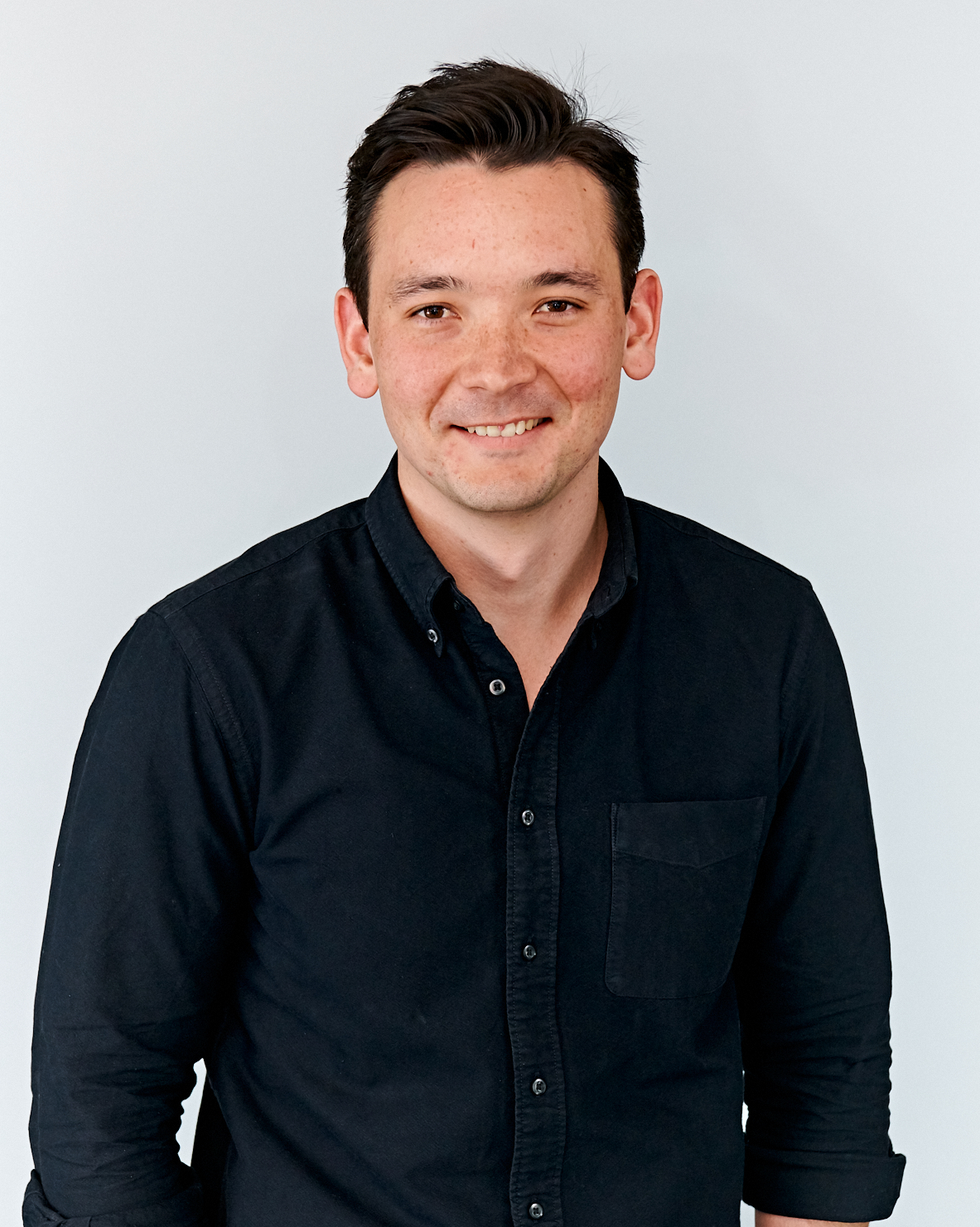}}]{Shaw Kudo} is a Mechanical Engineer at Arup and has worked on a broad range of projects as lead engineer and project manager. He has several years of experience in built environmental research and digital transformation, and is heavily involved in the development of Arup’s digital twin capabilities. His core research interests include exploring new ways of leveraging data to improve people’s lives, and digital strategy development.

\end{IEEEbiography}

\begin{IEEEbiography}[{\includegraphics[width=1in,height=1.25in,clip,keepaspectratio]{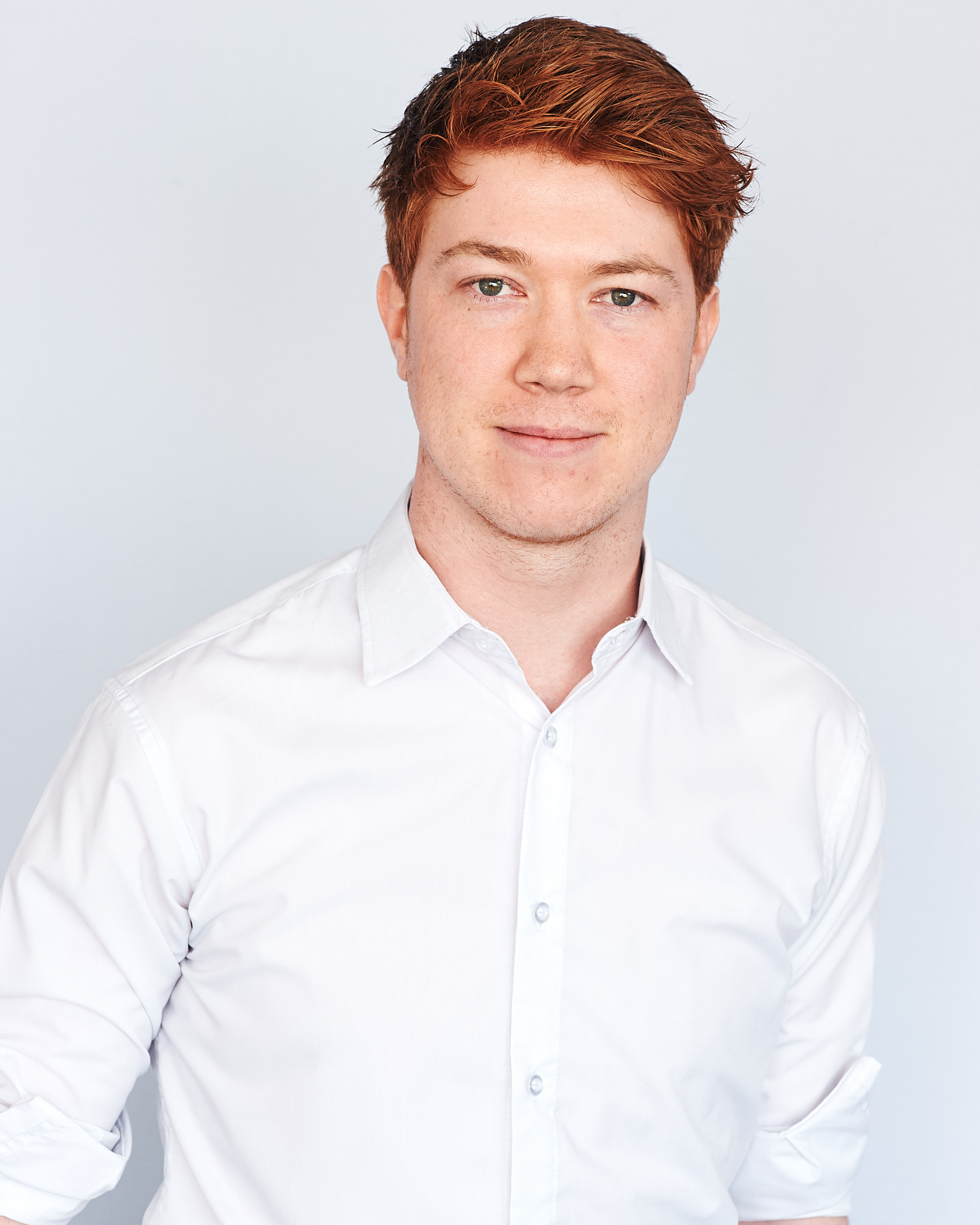}}]{Tim Rawling} began his engineering career at Arup Melbourne in 2016 and has worked on a broad range of commercial and charitable projects throughout Australasia. He is heavily involved in Arup’s development of building level digital twins as a product and service to clients in all sectors. His professional and research interest include data driven user centric building design, machine learning and strategic analysis.
\end{IEEEbiography}

\begin{IEEEbiography}[{\includegraphics[width=1in,height=1.25in,clip,keepaspectratio]{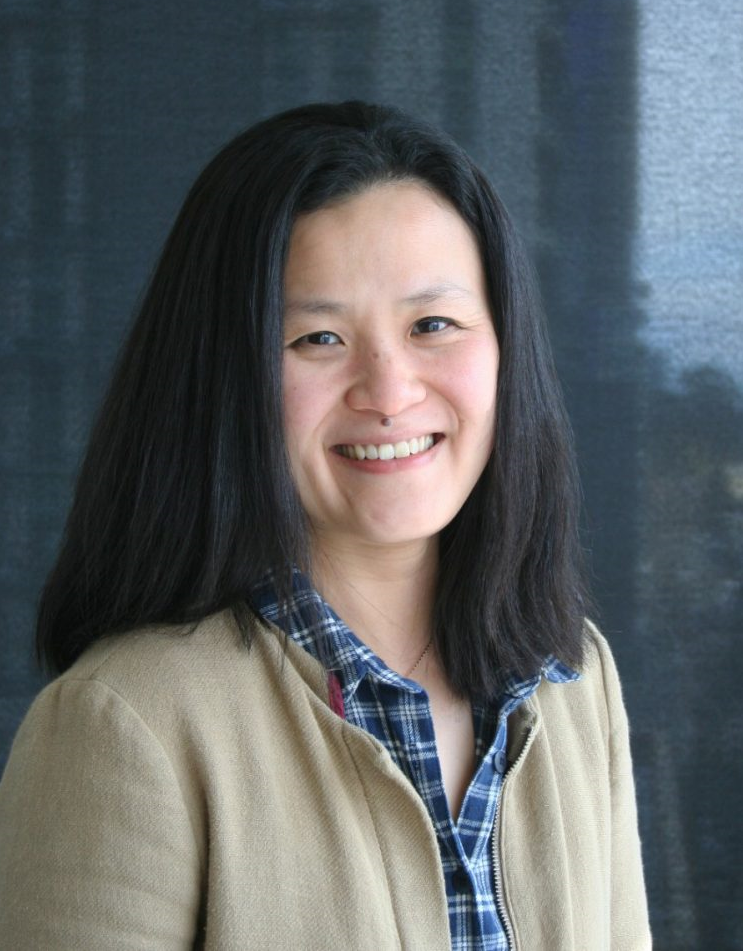}}]{Flora D. Salim}
received the Ph.D. degree from Monash University in 2009. She is currently an Associate professor of computer science and IT
with the School of Science, RMIT University. Her
research interests are human mobility and behavior analytics, context and activity recognition, and urban intelligence. Her research in spatio-temporal data analytics, context recognition and behavior
recognition, and prediction from sensor data has been evaluated across multiple projects, such as indoor monitoring and analytics in university and retail environments, driving behavior recognition, road risk analysis, and passenger movement analysis in airports. She is an Editorial Board Member of the Pervasive and Mobile Computing journal, an Expert Member of the International Energy Agency’s Energy in Buildings and Communities programme (Annex 79), and a Panel Member of JPI Urban Europe. She received the Australian Research Council Postdoctoral Fellowship Industry from 2012 to 2015. She was a recipient of the RMIT Vice-Chancellor’s Award for Research Excellence - Early Career Researcher 2016 and the RMIT Award for Research Impact - Technology 2018. She was a recipient of the Victoria Fellowship 2018 from the Victorian Government. She is also a Technical Program Committee Vice Chair of the 2018 IEEE PerCom. She is a Regular Invited Reviewer for the ACM Transactions On Internet Technology, Pervasive and Mobile Computing (Elsevier), the IEEE Transactions On Human-Machine Systems, the IEEE Transactions On Services Computing, the IEEE Transactions On Intelligent Transportation Systems, the IEEE Transactions On Cloud Computing, and the Data Mining and Knowledge Discovery journal.
\end{IEEEbiography}




\end{document}